\documentstyle[a4wide,12pt]{article}

\input psfig.tex

\begin{document}
\bibstyle{unsrt}

\begin{flushright}
Imperial/TP/94-95/20 \\
%hep-th/9502386
\end{flushright}

\begin{center}
{\huge Convergence of the Optimized Delta
Expansion for the Connected Vacuum Amplitude:
Anharmonic Oscillator\\}
\vspace{1cm}
{\Large C. Arvanitis, H. F. Jones and C. S. Parker \\}
\vspace{0.75cm}
{\large {\em Physics Department, Imperial College,\\
 London SW7 2BZ, United Kingdom.}}
\end{center}

\vspace{0.5cm}

\begin{abstract}
The convergence of the linear $\delta$ expansion for the connected
generating functional of the quantum anharmonic
oscillator is proved.  Using an order-dependent scaling for the
variational parameter $\lambda$, we show that the expansion converges
to the exact result with an error proportional to $\exp(-cN^{1/3})$.
\end{abstract}

\newpage
\section{Introduction}
\label{sec-intro}

 The linear delta expansion (LDE) is an analytic approach to field
theory which has been applied to a variety of problems (see
for example Ref.~\cite{jones_conf}).  The approach is non-perturbative
in the sense that a power series expansion is made in a parameter
$\delta$ artificially inserted into the action which
interpolates between a soluble action $S_0$ and the action for the
desired theory $S$.  The action is written
\begin{equation}
    S_{\delta}  =  (1 - \delta ) S_{0}  + \delta S,
    \label{eq-lde_def}
\end{equation}
where $S_0$ contains some dependence on a variational parameter
$\lambda$.  The generating functional for the theory may
be evaluated as a power series in $\delta$, which of course is usually
evaluated to finite order.  When $\delta$ is set equal to one at the
end of a truncated calculation of, say, a Green function $G$, some
dependence on $\lambda$ remains, which is where the variational
procedure makes its appearance.  One such procedure is the principle of
minimal sensitivity (PMS) \cite{pms} which requires $\lambda$ to be a
stationary point of the truncated Green function $G_N$:
\begin{equation}
    \frac{\partial G_N (\lambda)}{\partial \lambda} = 0   .
    \label{eq-pms}
\end{equation}

The PMS can provide both non-perturbative behaviour and convergence.
It has been shown to ensure the convergence of the
linear delta expansion for the vacuum generating functional for
$\phi^4$ theories in both zero \cite{buc_dunc_jones} and one
\cite{dunc_jones} dimension.
For the finite-temperature partition function $Z_N(\beta)$ at
odd $N$ there is only one stationary point, a global maximum, so
there is no ambiguity. Problems can, however, arise when there are
multiple solutions to the PMS condition Eq.~(\ref{eq-pms}), as is
the case \cite{ben_dunc_jones} for the connected vacuum generating
functional $W=\ln Z$ in zero dimensions.  There, rather than applying
the PMS directly to $W$, an order-dependent scaling for $\lambda$ was
chosen which guarantees convergence for $W$.
This was in fact the scaling resulting from the application of the PMS
to the calculation of $Z$.  In the present paper, we use a similar
choice of scaling to provide a convergent series of approximants for
the connected vacuum generating functional of the anharmonic oscillator.

The simplest choice of delta expansion used in Ref.~\cite{dunc_jones}
(using as trial action a free action with variable mass) gives the
finite-temperature partition function as a function of $z$, the complex
extension of
$\delta$:
\begin{equation}
	Z(z) = \frac{1}{Z_0} \int_{x(0)=x(\beta)} \, Dx \, \exp
	\left[- \int_0^\beta d\tau \,\left( \frac{1}{2} \dot{x}^2 +
	\frac{1}{2} (m^2 + 2g\lambda)x^2 + zg(x^4-\lambda x^2)\right)
	\right]
\label{eq-Zdef0}
\end{equation}
where $Z_0$ is the partition function for $g=0$.
It is this system that we consider here.  The scaling required for the
PMS which also gives convergence was shown in Ref.~\cite{dunc_jones}
to be $\lambda \propto N^{2/3}$.  In the present paper, we use this
scaling to prove the convergence of the $\delta$ expansion for
$W=\ln\;Z$,
provided that $m^2>0$.  This scaling is not strictly PMS, simply a
choice to give guaranteed convergence.

In the following Sec.~\ref{sec-method}, we review the proof of
convergence in the zero dimensional model.  This serves as a detailed
preview of the arguments used in the main sections of the paper.  In
Sec.~\ref{sec-rem}, we set out the asymptotic evaluation of the
remainder $R_N(z)=Z(z)-Z_N(z)$ for the anharmonic oscillator.
The analysis is closely related to that of Ref.~\cite{dunc_jones}.  In
Sec.~\ref{sec-sadd} we discuss the evaluation of
Eq.~(\ref{eq-Zdef0}) for $Z(z)$ by saddle point methods and the
cancellations that
govern the convergence of $W_N$. We compare our analysis with numerical
calculations in Sec.~\ref{sec-num}, and summarize our results in the
conclusions, Sec.~\ref{sec-conc}.

\section{Illustration of the method}
\label{sec-method}
The method of the present paper is similar in spirit to that of
Ref.~\cite{ben_dunc_jones}, which considers the zero dimensional
analogue of the vacuum generating functional, namely:
\begin{equation}
	Z^{[0]}(z)=\int_{-\infty}^{\infty} dx \,\exp(-[\lambda (1-z)
	x^2 +gzx^4])
\end{equation}
An expansion of $Z^{[0]}$ in powers of $z$ was proved in
Ref.~\cite{buc_dunc_jones} to converge to the correct value for
$Z^{[0]}(1)$ with an error like $e^{-cN}$.  The PMS condition guarantees
convergence and requires $\lambda$ to scale as $\sqrt{N}$.

In Ref.~\cite{ben_dunc_jones} it was shown that the convergence of
$W_N=\{\ln Z_N(z)\}_N|_{z=1}$, where $\{f(z)\}_N$ denotes the Taylor
expansion of $f(z)$ truncated at $O(z^N)$, is governed by the position
of the roots of $Z_N(z)$.  If any of these roots lies inside the
circle $| z|\leq1$, the expansion for $W_N(z)$ is not convergent,
and in fact the position of the smallest modulus root determines the
rate of
convergence of the series for $W_N$.  The difficulty in such a proof
is that while it is a relatively simple matter to calculate the
coefficients $c_n$ in the series $Z_N(z)=\Sigma c_n z^n$ (using simple
integrals in zero dimensions or the methods of Caswell
\cite{caswell} and Killingbeck \cite{kill} in one dimension), finding
the roots analytically is highly non-trivial.

In order to circumvent this difficulty, we write
\begin{equation}
	Z_N(z) = Z(z) - R_N(z)
\label{eq-Rdef}
\end{equation}
{}From this point of view, zeros of $Z_N(z)$ arise from a cancellation
between $Z(z)$ and $R_N(z)$, each of which can be evaluated
by asymptotic methods.  In
Ref.~\cite{ben_dunc_jones}, the remainder $R_N(z)$ is bounded quite
simply, while $Z(z)$ is estimated by stationary phase integration.

The saddle points of the exponent $S$ in Eq.~(4) are $x_0 = 0$ and
$x_\pm = \pm \lambda (1-z)/2gz$. The mechanism identified in
Ref.~\cite{ben_dunc_jones} for
roots to lie inside the circle of mean radius $|z| = |c_0/c_N|^{1/N}$ was
cancellation between the saddle-point contributions from $x_0$ and
$x_\pm$. This can only occur along the Stokes line $\Gamma$ defined by
${\rm Re}\;S(x_\pm) = 0$. However, part of $\Gamma$ lies inside $|z|=1$,
and cancellation along this part of the curve would have been
disastrous for the proof of convergence. Such cancellations were ruled
out in Ref.~\cite{ben_dunc_jones} by explicitly constructing the
stationary phase paths
for different values of $z$. It was shown that for $|z|<1$ the correct
path did not pass through $x_\pm$, in contrast to the region $|z|>1$,
where all three saddle points were traversed.

Because this argument is difficult to generalize to a genuine path
integral, we note here a simpler argument ruling out dominant
contributions from $x_\pm$ in the region ${\rm Re}\;z <1$. In fact
$Z^{[0]}(z)$ may be evaluated exactly as a modified Bessel function
$K_{1 \over 4}((\lambda(1-z))^2/(8g^2z^2))$, provided that ${\rm Re}\;z
<1$. The asymptotic behaviour of this function as $\lambda \rightarrow
\infty$ has no exponential dependence on $N$ in this region. Thus
saddle points potentially giving a greater contribution than the
Gaussian saddle point $x_0$ are excluded. The argument immediately
generalizes to the left half plane without the need for
contour rotation.  The analysis in the present paper is similar,
though we are unable to escape the need for contour rotation.

In the region ${\rm Re}\;z>1$ all three saddle points can occur, with
possible cancellations along the curve $\Gamma$. In this way it was
shown that the smallest zero of $Z^{[0]}$ occurs at
 $z_{\rm min}= a + \sqrt{a^2-1}$
where $a=1+3 \pi i /(N \alpha)$, with $\alpha=1.3254...$.  This tends
to unity sufficiently slowly as $N$ increases to ensure that
$W_N^{[0]}$ converges to $W^{[0]}$ with an error of the order of
$\exp(-\sqrt{3 \pi N /\alpha})$.

\section{Convergence of the remainder $R_N(z)$}
\label{sec-rem}

The remainder $R_N = Z(z)-Z_N(z)$ can be written as
\begin{eqnarray}
\label{eq-rem}
	R_N(z) & = & \frac{1}{Z_0} \int_{x(0)=x(\beta)} \, Dx \, \exp
	\left[-\int_0^\beta d\tau \,\left( \frac{1}{2} \dot{x}^2 +
	\frac{1}{2} (m^2 + 2g\lambda)x^2 + zg(x^4-\lambda
	x^2)\right)\right] \nonumber\\ & & \times \, [1-\Theta_N(y)]\;,
\end{eqnarray}
where $\Theta_N(y) := e^{-y} \{e^y\}_N$ and
$y = -gz\int\;d\tau (x^4-\lambda x^2)$.

By integrating the identity for $\Theta_N$ given in
Ref.~\cite{buc_dunc_jones}:
\begin{equation}
\label{eq-theta}
	\frac{d}{dy} \Theta_N = - \frac{ y^N e^{-y}}{N!} \, ,
\end{equation}
we can write
\begin{equation}
	1 - \Theta_N(y) = \int_0^y \, dy^\prime \, \frac{y^{\prime N}
    	e^{-y^\prime }}{N!} .
\end{equation}
Since $z$ and hence $y$ is in general complex, we must make a change
of variable dependent on the sign of the real part of $y$.  For
${\rm Re}\;y \ge 0 $ , we may write $y=|y|e^{i\theta}$ and change
variables using $y^\prime = \omega e^{i\theta}$
\begin{equation}
	1 - \Theta_N(y) = \int_0^{|y|} \, d\omega \,
	\frac{\omega^N}{N!} \exp \left(-\omega e^{i\theta} +
	i(N+1)\theta\right)          .
\end{equation}

For ${\rm Re}\;y \le 0$, we write $y=|y|e^{i(\theta +\pi)}$ ($-\pi/2 \le
\theta \le \pi/2$) and use $y^\prime = \omega e^{i(\theta +\pi)}$:
\begin{equation}
	1 - \Theta_N(y) = \int_0^{|y|} \, d\omega \,
	\frac{\omega^N}{N!} \exp\left(-\omega e^{i\theta} +
	i(N+1)(\theta+\pi)\right)
\end{equation}

Applying these to the remainder Eq.~(\ref{eq-rem}) for ${\rm Re}\;z \ge
0$, we define weak and strong field regimes respectively by
\begin{eqnarray}
    	\lambda \int_0^\beta \, d\tau \, x^2 & \ge & \int_0^\beta \,
	d\tau \, x^4         \nonumber \\
	\lambda \int_0^\beta \, d\tau \, x^2 & \le & \int_0^\beta \,
	d\tau \, x^4         .
\end{eqnarray}
We divide the remainder $R_N(z)$ into weak and strong field
contributions $A_N$ and $B_N$, and find
\begin{eqnarray}
\label{eq-An_def}
	|A_N(z)| & = & \frac{1}{Z_0} \int_A \, Dx \, \exp\left[-\int
	\, d\tau \, \left(\frac{1}{2} \dot{x}^2 +
	\frac{1}{2}(m^2+2g\lambda)x^2 +  g|z|\cos\theta \, (x^4 -
	\lambda x^2)\right)\right] \nonumber \\
    	& & \times \,\int_0^{|y|} \, d\omega \, \frac{\omega^N}{N!}
	\exp\left(-\omega \cos\theta\right) \\
\label{eq-Bn-def}
    	|B_N(z)| & = & \frac{1}{Z_0} \int_B \, Dx \, \exp\left[-\int
	\, d\tau \, \left(\frac{1}{2} \dot{x}^2 + \frac{1}{2}
	(m^2+2g\lambda)x^2 + g|z|\cos\theta \, (x^4 - \lambda x^2)
	\right)\right] \nonumber \\
    & & \times \,\int_0^{|y|} \, d\omega \, \frac{\omega^N}{N!}
	\exp(\omega \cos \theta)  .
\end{eqnarray}

The two regimes are dealt with separately in the following
subsections.  For ${\rm Re}\;z<0$, we simply interchange the role of the
weak and strong field contributions to $R_N(z)$.  The calculations
give a result similar to the zero dimensional case, namely that we may
bound the remainder by
\begin{equation}
	|R_N(z)| \le D |z|^{N+1}  \exp\left(-cN^{2/3}\right).
\label{eq-r_n}
\end{equation}
The detailed derivation of Eq.~(\ref{eq-r_n}) for $|z|=1$ and its
reliance on the
PMS scaling $\lambda = (2\gamma^2 N^2/g)^{1/3}$ with $\gamma=0.186$
are given in Ref.~\cite{dunc_jones}.  For $|z| \neq 1$ we see that
$R_N(z)$ is convergent for
\begin{equation}
	|z| < 1 + O(N^{-1/3})\;.
\end{equation}

\subsection{Strong fields}
This is the simpler regime to consider, as the analysis almost
exactly mirrors that of Ref.~\cite{dunc_jones}.  Since the $\omega$
integrand in Eq.~(\ref{eq-Bn-def}) is monotonically increasing, we
may simply bound by the value at the upper limit:
\begin{eqnarray}
	|B_N(z)| & \le & \frac{1}{Z_0 \, N!} \int_B \, Dx \, \exp
	\left[-\int \, d\tau \, \left(\frac{1}{2}
	\dot{x}^2+\frac{1}{2}(m^2+2g\lambda)x^2
    	+ g|z|\cos\theta(x^4 - \lambda x^2)\right)\right] \nonumber \\
    	& & \times |y|^{N+1}\exp\left(|y|\cos\theta\right) .
\end{eqnarray}
Now using $|y| = g|z| \int \, d\tau \, [x^4 - \lambda x^2]$, we
find that
\begin{eqnarray}
    	|B_N(z)| & \le & \frac{|z|^{N+1}}{Z_0 N!} \int_B \, Dx \, \exp
	\left[-\int \, d\tau \, \left(\frac{1}{2} \dot{x}^2 +
	\frac{1}{2}(m^2 +2g \lambda)x^2\right)\right] \nonumber\\
    	& & \times \left(g \int \, d\tau \, \left(x^4 - \lambda
	x^2\right)\right)^{N+1}\;,
\end{eqnarray}
which is just $|z|^{N+1}$ times the $B_N$ given in
Ref.~\cite{dunc_jones}.  Thus with no further analysis we are able to
bound the strong field remainder by
\begin{equation}
	|B_N(z)| \le |z|^{N+1} C(m,g) \beta N^{4/3}
	\exp\left(-NS_B(\gamma)\right)
\end{equation}
with $S_B$ and $\gamma$ as defined in Ref.~\cite{dunc_jones}.

\subsection{Weak fields}
The weak field contribution to the remainder is given by
\begin{eqnarray}
	|A_N(z)| & = & \frac{1}{Z_0} \int_A \, Dx \, \exp\left[-\int
	\, d\tau \, \left(\frac{1}{2} \dot{x}^2 +
	\frac{1}{2}(m^2+2g\lambda)x^2 +  g|z|\cos\theta(x^4 -
	\lambda x^2)\right)\right] \nonumber \\
    	& & \times \,\int_0^{|y|} \, d\omega \, \frac{\omega^N}{N!}
	\exp\left(-\omega \cos\theta\right) .
\end{eqnarray}
If we make a further change of variable $\omega = g|z|\sigma \int
d\tau [\lambda x^2 - x^4]$, the familiar factor $|z|^{N+1}$ already emerges:
\begin{eqnarray}
	|A_N(z)|  & = & \frac{|z|^{N+1}}{Z_0 N!} \int_0^1 \,d\sigma
	\,\int_A \, Dx \, \exp\left[-\int \, d\tau \,
	\left(\frac{1}{2} \dot{x}^2 +
	\frac{1}{2}(m^2+2g\lambda)x^2\right)\right]  \nonumber \\
	& & \times \, \sigma^N \left( g \int \, d\tau \, (\lambda x^2- x^4)
	\right)^{N+1}  \exp \left[g|z|\cos\theta(1-\sigma)\int\, d\tau
	\, (\lambda x^2 - x^4)\right]
\end{eqnarray}
For the allowed range of $\theta$, it is then clear that
\begin{eqnarray}
\label{eq-An3}
	|A_N(z)|  & \le & \frac{|z|^{N+1}}{Z_0 N!} \int_0^1 \,d\sigma
	\,\int_A \, Dx \, \exp\left[-\int \, d\tau \,
	\left(\frac{1}{2} \dot{x}^2 +
	\frac{1}{2}(m^2+2g\lambda)x^2\right)\right]  \nonumber \\
	& & \times \, \sigma^N \left( g \int \, d\tau \, (\lambda x^2- x^4)
	\right)^{N+1}  \exp \left[g|z|(1-\sigma)\int\, d\tau
	\, (\lambda x^2 - x^4)\right]
\end{eqnarray}

We may then further bound $A_N(z)$ by removing the term
$(1-\sigma)gzx^4$ from the last exponent in Eq.~(\ref{eq-An3}).  We
also use Stirling's approximation for the factorial.  We follow
Ref.~\cite{dunc_jones} and allow the calculation to proceed for either
sign of $m^2$ by writing:
\begin{eqnarray}
\label{eq-An4}
	|A_N| & < & \frac{|z|^{N+1}}{\sqrt{2 \pi N} Z_0}\int_0^1\,
	d\sigma \int_A \,Dx\left(g \,\int \, d\tau \left(\lambda
	x^2-x^4\right)\right) \exp \left[ - \int _0^\beta \, d\tau
	\left(\frac{1}{2} \dot{x}^2 + \frac{1}{2}
	|m^2|x^2\right)\right] \nonumber\\ & & \times \exp \left(-N
	S_A(x, \sigma, z) \right)
\end{eqnarray}
where
\begin{eqnarray}
\label{eq-Sa-def}
	S_A(x,\sigma,z) & =& \int_0^\beta \, d\tau \,
	\left(\frac{x^2}{N}\left(g\lambda\left(1-|z|(1-\sigma)\right)
	+ m^2 \theta(-m^2)\right)\right) \nonumber \\
	& & - \ln \left(\frac{g\sigma}{N} \int_0^\beta \, d\tau \,
	\left(\lambda x^2 - x^4 \right)\right) -1\;,
\end{eqnarray}
which for $z = 1$  is equal to the $S_A$ defined in
Ref.~\cite{dunc_jones}.

The calculation proceeds in exactly the same manner as in
Ref.~\cite{dunc_jones}, with the definition $\int_0^\beta \, x^2 \,d\tau
= \beta \lambda U$ and the use of the Cauchy--Schwarz identity to show
that
\begin{equation}
\label{eq-Sa2}
	S_A \ge \tilde{\alpha} U - \ln \left(\alpha U(1-U)\right) -1\;,
\end{equation}
with $\alpha$ and $\tilde{\alpha}$ the $z \ne 1$ extensions of those
defined in Ref.~\cite{dunc_jones}:
\begin{eqnarray}
\label{eq-alphadef}
	\alpha & = & \frac{g \beta \lambda^2 \sigma}{N}
	\nonumber \\
	\tilde{\alpha} & = &  \alpha\left(|z|+\frac{1-|z|}{\sigma}\right)
	+ \frac{m^2\beta \lambda}{N} \theta(-m^2) .
\end{eqnarray}
We minimize the bound on $S_A$ with respect to $U$ and take the minimum at
\begin{equation}
	U = \frac{1}{2} + \frac{1}{\tilde{\alpha}} \left[ 1- \sqrt{1 +
	\frac{\tilde{\alpha}^2}{4}}\,\right],
\end{equation}
which gives a similar result to that of Ref.~\cite{dunc_jones}:
\begin{equation}
	S_A \ge F(\tilde{\alpha}) - \ln \left(\frac{\alpha}{2}\right)\;,
\end{equation}
where
\begin{equation}
	F(\tilde{\alpha}) = \frac{\tilde{\alpha}}{2} - \sqrt{1 +
	\frac{\tilde{\alpha}^2}{4}} + \ln\left(1+ \sqrt{1 +
	\frac{\tilde{\alpha}^2}{4}}\right)  .
\end{equation}
It is true for the present  definition of $\tilde{\alpha}$ that for
$\lambda \propto N^{2/3}$ and large $N$,
\begin{equation}
	\frac{\partial}{\partial \alpha}\left [F(\tilde{\alpha}) - \ln
	\left(\frac{\alpha}{2}\right)\right] < 0 \;,
\end{equation}
so that the integral over $\sigma$ in Eq.~(\ref{eq-An4}) is
dominated by its value at the upper limit.  Thus:
\begin{eqnarray}
\label{eq-An5}
	|A_N| & < & \frac{|z|^{N+1}}{\sqrt{2 \pi N}Z_0} \int_A \, Dx
	\left(g \int	\, d\tau \left[\lambda x^2 - x^4\right]\right)
	\, \exp \left[ - \int_0^\beta \, d\tau \, \left[\frac{1}{2}
	\dot{x}^2 + \frac{1}{2}|m|^2 x^2\right]\right]  \nonumber \\ &
	& \times \exp \left[ - N\left(F(\tilde{\alpha}_0) -
	\ln\left(\frac{1}{2}\alpha_0\right)\right)\right]
\end{eqnarray}
with the subscript 0 on $\alpha$ and $\tilde\alpha$ denoting
$\sigma = 1$, giving
\begin{eqnarray}
	\alpha_0 & = & \frac{g \beta \lambda^2}{N} \nonumber \\
	\tilde{\alpha}_0 & = & \alpha_0 + \frac{m^2 \beta \lambda}{N}
	\theta(-m^2) .
\end{eqnarray}
Now if we use the relation: $g \int _0^\beta d\tau \, [\lambda x^2 -
x^4] \le g \beta \lambda^2 /4$ and the fact that the functional
integral in the numerator of Eq.~(\ref{eq-An5}) is bounded above by that
in the denominator, we obtain
\begin{equation}
	|A_N| \le \frac{g\beta \lambda^2 |z|^{N+1}}{4 \sqrt{2 \pi N}}
	\exp \left[ -N \left(F(\tilde{\alpha}_0) -
	\ln\left(\frac{\alpha_0}{2}\right) \right) \right]\;.
\end{equation}

Using the scaling $\lambda \propto N^{2/3}$ we have $\tilde{\alpha}_0
\approx \alpha_0 [ 1 + O(N^{-2/3})]$.  For large $N$, $\alpha \propto
N^{1/3}$ and we may finally write
\begin{equation}
	|A_N(z)| \le |z|^{N+1}N^{5/6} \exp\left[-c
	N^{2/3}\right] .
\end{equation}

\section{Saddle point expansion of $Z(z)$}
\label{sec-sadd}
The partition function may be written as
\begin{equation}
	Z(z) = \frac{1}{Z_0}\int_{x(0)=x(\beta)} \, Dx \exp
	\left(-S[x] \right)
\label{eq-Zdef}
\end{equation}
where
\begin{equation}
	S[x] := \int _0^\beta \, d\tau \, \left( \frac{1}{2}
	\dot{x}^2 + \frac{1}{2} M^2 x^2 +gzx^4\right)
\label{eq-Sdef}
\end{equation}
and $M^2 = m^2 + 2g\lambda (1-z)$, which is large for $\lambda
 \propto N^{2/3}$ and $z \neq 1$.

Performing a saddle point expansion of the path integral requires
finding the solutions of the Euclidean equation of motion
\begin{equation}
\label{eq-mot}
	\ddot{x}  =  M^2 x + 4 g z x^3 \;,
\end{equation}
with $x(0) = x(\beta)$, which has the static solutions:
\begin{eqnarray}
\label{eq-x0}
	x =& x_0 & =  0 \\
\label{eq-xpm}
	x= & x_{\pm} & =  \pm \sqrt{-\frac{M^2}{4gz}}
\end{eqnarray}

There is also a non-static instanton solution.  This is discussed by
Zinn-Justin \cite{zinn} in the case of negative $g$ and real $M^2>0$.
This class of solution also arises in the strong coupling
contribution to the remainder in Ref.~\cite{dunc_jones}, the modified
action giving an equation of motion equivalent to {\em negative
coupling}.  This solution is
\begin{equation}
\label{eq-inst}
	x_I(\tau) = \sqrt{\frac{-M^2}{2gz}}\frac{1}{\cosh(M(\tau -
	\tau_0))}  ,
\end{equation}
where $\tau_0$ must strictly be taken as $\tau_0 = {1 \over 2}\beta$ to
satisfy the periodic boundary conditions. We note that this solution takes
the form
$1/\cos(M(\tau -\tau_0))$ for real $z$, such that $z>1+m^2/(2g
\lambda)$.  For general complex $z$ the solution is an hybrid.

A more general class of solutions is discussed by Richard and Rouet
\cite{Rich_Rou} for the {\em double well} oscillator.  The action of
Eq.~(\ref{eq-Sdef}) may be rescaled to the form of Ref.~\cite{Rich_Rou}
which is:
\begin{equation}
	S_{RR}[x] = \int_{-T}^T dt \, \left[ \frac{1}{2} \dot{x}^2
	+ \frac{1}{2} ( x^2 -1)^2\right]
\end{equation}
The solutions to the modified equation of motion take the form
\begin{equation}
	x_{RR}(t) = a \left(1 + \frac{a^2-1}{{\cal P}(t +
	u |\omega,\omega^\prime)+\frac{1}{6}
	- {1 \over 2}a^2}\right)
\label{eq-RR}
\end{equation}
where $a$ is an integration constant and $ {\cal P}(t +
u|\omega,\omega^\prime)$ is a Weierstrass elliptic function
\cite{ellip} whose periods $\omega,\omega^\prime$ are determined by
the boundary conditions and form of the elliptic integral inverted by
Eq.~(\ref{eq-RR}).  The cosine instanton solution
discussed above is equivalent to Eq.~(\ref{eq-RR}) with $\omega^\prime
/ \omega \rightarrow i\infty$.

The classical action for all the saddle points at large $M$
may be evaluated simply:
\begin{eqnarray}
	S[x_0] & = & 0 \label{eq-0act}\\
	S[x_\pm] & = & -\frac{M^4 \beta}{16gz} \label{eq-pmact}\\
	S[x_I] & = & -\frac{M^3}{3gz} \label{eq-instact}\\
	S[x_{RR}] & = & S[x_\pm] + k\,S[x_I] \label{eq-rract}\;,
\label{eq-acts}
\end{eqnarray}
where $k$ is a positive integer, except that in the case
$\omega^\prime/\omega\rightarrow i\infty$,
cancellations occur in the action of Richard and Rouet to give
$S[x_{RR}] = S[x_I]$.

Having evaluated the classical actions, we must evaluate the regions
of the $z$ plane in which the various saddle points are dominant.
These regions are illustrated in Fig.~1 for the numerical
values $m=1$, $g=1/2$, $\beta=2$ and $\lambda = 11.619$ ($N=75$).
The solid curve $\Gamma$ is the curve ${\rm Re}\;S[x_\pm] = 0$ and
the dotted curve $\Gamma_I$ is the curve ${\rm Re}\;S[x_I] = 0$,
while the dashed vertical line marks ${\rm Re}\;z=1+m^2/2g\lambda$.
All three meet at the point $z=1+m^2/2g\lambda$.

Inside $\Gamma_I$, the Richard and Rouet saddle points are dominant.
Inside the loop of $\Gamma$, the saddle point $x_{\pm}$ dominates
over $x_0$, though this is unimportant given the overall dominance
of $S[x_{RR}]$. Outside $\Gamma_I$
the zero saddle point becomes dominant over all its competitors.  This
remains the case up to the boundary $\Gamma$ to region C, in
which $x_{\pm}$ is dominant.  Neither $x_I$ nor $x_{RR}$ can
dominate the integral in this region.

It would then appear that instantons dominate for small $|z|$,
$x_0$ is dominant for intermediate $|z|$, and for large $|z|$ the
saddle points $x_{\pm}$ are the important ones. However, as we learnt
from the zero-dimensional case, it is important to check whether
apparently dominant saddle points do in fact contribute. In zero dimensions
Ref.~\cite{ben_dunc_jones} it was possible to trace out the
stationary-phase paths and to see that these did not pass through the
non-zero saddle points when $|z|<1$.  For a
path integral, such a procedure is extremely ill defined. However, we
are able to use instead an analogue of the Bessel function analysis
given
above for the zero dimensional case by realizing that for ${\rm Re}\;z <
1+m^2/(2g\lambda)$  the action is generically a real single well, the
complex part of the mass term simply adding a phase.  Provided then
that ${\rm Re}(gzx^4) >0$ the partition function is bounded above by that
for the pure harmonic oscillator.  The harmonic oscillator partition
function is given by
\begin{eqnarray}
	|Z_0| & = & \int Dx \, \exp \left[- \int_0 ^\beta d\tau\,
	\left( \frac{1}{2} \dot{x}^2 + \frac{1}{2} {\rm Re}(M^2) x^2
	\right) \right] \nonumber \\
	& = & C \big({\det}(\partial_t^2 + {\rm Re}M^2)\big)^{-\frac{1}{2}}\;,
\end{eqnarray}
where $C$ is a normalization constant.  The Gaussian integration about the
saddle point $x=0$ in Eq.~(\ref{eq-Zdef}) gives:
\begin{equation}
	|Z| = C | {\det}(\partial_t^2 + M^2)|^{-\frac{1}{2}}\;.
\end{equation}
We are thus able to state that in the region $0<{\rm Re}\;z<1+m^2/(2g\lambda)$
there is in fact no exponentially increasing contribution from saddle points
with ${\rm Re}\;S[x]<0$.  The zero saddle point is the only one which can
contribute.  From the form of Eq.~(\ref{eq-r_n}), it is then clear that
no zeros of $Z_N(z)$ are possible for $|z|<1+O(N^{-1/3})$ in the region
$0<{\rm Re}\;z<1+m^2/(2g\lambda)$. In fact the zeros will be expected to
occur on a ring of radius $1+O(N^{-1/3})$.

The left half of the $z$ plane may be included by rotating the
contour of integration to make ${\rm Re}(zx^4)>0$ while keeping both
${\rm Re}(\dot{x}^2) >0$ and ${\rm Re}(M^2 x^2)>0$.  This procedure is similar
to that discussed in Ref.~\cite{zinn}.  In the upper half plane, we
rotate the contour by an angle $\phi = (\pi/2 -\theta)/4$, where
$\theta = \arg z$.  In the lower half plane, we simply rotate the
contour by an equivalent amount, but in the opposite sense.

In the region ${\rm Re}\;z>1 +m^2/(2g\lambda)$, we need only consider the
stationary saddle points $x_0$ and $x_{\pm}$.  In the region B, $x_0$
dominates and the requirements for cancellation are the same as
those discussed above.  No zeros affecting the rate of convergence
of $W_N$ can arise in this region for $m^2>0$, and we expect the ring of
zeros to continue smoothly into this region.  For the non-Borel summable
case $m^2<0$, this
proof breaks down, since zeros of $Z_N(z)$ could now occur for
$|z|\le 1$.

In region C, where $x_\pm$ is the dominant saddle point,
$Z(z)$ contains exponential dependence on $N$, allowing
cancellation with $R_N(z)$, which grows with $|z|$ for
$|z|>1+O(N^{1/3})$. The zeros in region C will thus tend to be further
out than those in B.

An exception to this is the region close to $\Gamma$, the Stokes line
where ${\rm Re}\;S[x_{\pm}]=0$.
There cancellations can occur between the two saddle-point contributions
from $x_0$ and $x_\pm$, giving a smaller effective exponent for $Z(z)$, and
the possibility of roots of smaller modulus. However, since this branch of
$\Gamma$ lies to
the right of ${\rm Re}\;z=1+m^2/(2g\lambda)$, such zeros again do not
affect the convergence of $W_N$ for $m^2 > 0$. Having established that all the
roots satisfy $|z-1| > m^2/2g\lambda = O(1/N^{2/3})$, the
analysis of Ref.~\cite{ben_dunc_jones} shows us that $W_N$ tends to
the correct value $W$ with an error
\begin{eqnarray}
	{\cal R}_N & := & W-W_N \nonumber \\
	& \propto &  \exp \left(-c N^{1/3}\right)
\end{eqnarray}

\section{Comparison with numerical results}
\label{sec-num}
The above analysis of the position of the roots of $Z_N(z)$ can be checked
numerically.
Figs.~2, 3 and 4 show the
roots of $Z_N(z)$ for $\beta=2$, $m=1$, $g=1/2$ and $N=25$, $45$ and
$75$ respectively.  In each case, $\lambda$ is chosen as the
PMS value for $Z_N(z)$.  In the previous section, it was
noted that the smallest root occurs where ${\rm Re}\;S[x_{\pm}]=0$.  In
order to make the comparison, we have plotted the curve $\Gamma$ on
each graph.

In Fig.~2 we see that at order $25$, the roots lie to a
good approximation on a ring, the radius of which is determined by the
ratio of the
first and last coefficients in the $\delta$ series.
Fig.~3, for order $45$, shows that two pairs of roots have
broken away from this ring, one pair moving out and one pair moving
in.  The inner pair of roots have the smallest modulus, and lie
sufficiently close to $\Gamma$ to indicate that the mechanism outlined
above is indeed operative. At order $75$ in
Fig.~4, two pairs of roots are tracking in along the curve
$\Gamma$.  This
behaviour very closely follows that described in
Ref.~\cite{ben_dunc_jones} for the zero dimensional case.

We have checked the convergence of the expansion for $W$ numerically.
Taking $\lambda_N$ as the unique PMS value for $Z_N$ we obtain Table 1,
showing convergence to 15 significant figures by $N=85$ for both $\beta=2$
 and $\beta=5$. The convergence is not monotonic, as was the case for
$Z_N$; rather it oscillates around the exact value with exponentially
decreasing amplitude.

Notwithstanding the problem of multiple PMS points for $W_N(\lambda)$
itself, it is interesting to explore the dependence on $\lambda$ of
$W_N$. For the values of $N$ we have considered there seems to be an
initial, extremely broad maximum, which gives a very accurate estimate
of $W$, followed by a series of secondary
maxima and minima of progressively decreasing accuracy.
For $N=45$ and $\beta=2$ the value at this first maximum is
-1.353868180362611, differing from the exact value only in the 16th.
significant figure. A similar situation occurs for $\beta=5$.

\begin{table}[b]
\begin{tabular}{c c c}

$N$  & 		$W_N(\beta=2)$ 	    &	$W_N(\beta=5)$ \\
\hline

  17 & 		-1.3538682048367371 &  	-3.4805935670797027 \\
  25 & 		-1.3538681804666174 &  	-3.4805880725830912 \\
  35 &		-1.3538681803659796 &  	-3.4805878403631746 \\
  45 & 		-1.3538681803626513 &  	-3.4805878324251261 \\
  55 & 		-1.3538681803625918 &  	-3.4805878320234629 \\
  65 &		-1.3538681803626148 &  	-3.4805878319968700 \\
  75 & 		-1.3538681803626174 &  	-3.4805878319947731 \\
  85 &		-1.3538681803626168 &  	-3.4805878319945885 \\
\hline
Exact value & 	-1.3538681803626170 & 	-3.4805878319945603 \\
\end{tabular}
\caption{Results showing convergence of $\delta$ expansion for $W_N$}
\end{table}

\section{Conclusions}
\label{sec-conc}
In this paper, we have proved the convergence of the optimized
$\delta$ expansion for $W=\ln Z$, where $Z$ is the finite-temperature
partition function of the anharmonic oscillator.  With the variational
parameter $\lambda$ chosen to scale with the order $N$ of the expansion as
\begin{equation}
	\lambda = (2\gamma^2 N^2/g)^{1/3}
\end{equation}
we found that
\begin{equation}
	W-W_N \propto \exp \left(-c N^{1/3}\right).
	\label{eq-err}
\end{equation}

Apart from the fact that in a field theoretic context it is generally the
connected
Green functions which are the relevant quantities, the original motivation
\cite{ben_dunc_jones} for looking at $W$ was the non-uniform convergence with
$\beta$ of the delta expansion for $Z_N$, in particular in the limit
$\beta \rightarrow \infty$. Since this non-uniformity corresponds to the
limit of large spacetime volume, it seemed plausible that the difficulty
would be much less severe for $W$, the generating function of connected
diagrams, which should depend only linearly on the (large) volume cut-off.
However, the present proof of convergence for $W_N$ does not address this
problem, since it relies as an intermediate step on the convergence of the
sequence $Z_N$. It remains an open question whether the PMS criterion
applied directly to $W_N$ itself, in spite of the
problems of multiple stationary points, can give rise to uniform
convergence as $\beta\rightarrow\infty$.

In the meantime a paper by Guida, Konishi and Suzuki~\cite{guida} has
appeared, which, using a completely different approach involving a
dispersion relation in $g$, has proved convergence of the $\delta$ expansion
for the individual energy levels of the anharmonic oscillator for a variety of
scaling exponents $\eta$ in $\lambda_N \propto N^\eta$. The PMS choice
we have used here, with $\eta=2/3$, lies at the edge of their range
$2/3 \le \eta < 1$. It is possible that their method could be extended
to a proof of convergence for $W$, or the free energy $F=W/\beta$.

The proof of Ref.~\cite{guida} is restricted to $m^2>0$, as is the
case here, even though it was shown in Ref.~\cite{dunc_jones} that
for $Z$ the sequence of approximants $Z_N$ converges for either sign
of $m^2$. As a general principle the field theory about which we expand
should capture as closely as possible the essential features of the system
under investigation~\cite{dyson}. For the double-well oscillator it may well
be that to obtain a convergent expansion for the connected
Green functions a more sophisticated trial
action is needed than the one used here and in Ref.~\cite{guida},
namely a free action with positive $m^2$.

The extension of the method to higher dimensions  should be
possible, since only saddle point techniques are used.  In higher
dimensional field theories we must, however, take account of the
interplay between the $\delta$ expansion and the renormalization
procedure.  Such a scheme has been successfully applied to the
Gross-Neveu model in the large $N$ limit using the $\delta$ expansion
\cite{jones_moshe} and a related scheme \cite{carv} and $\phi^4$
theory in four dimensions in the Gaussian approximation
\cite{stan_stev}.

\section*{Acknowledgements}
C.~A. is grateful to the theory group at Imperial College for their kind
hospitality and acknowledges financial support from ECC grant no.
ERBCHBICT941235.

%%%%%%%%%%%%%%%%%%%%%%%%%%%%%%%%%%%%%%%%%%%%%%%%%%%%%%%%%%%%%%%%%%%%%%%

\begin{figure}
	{\centerline{\psfig{file=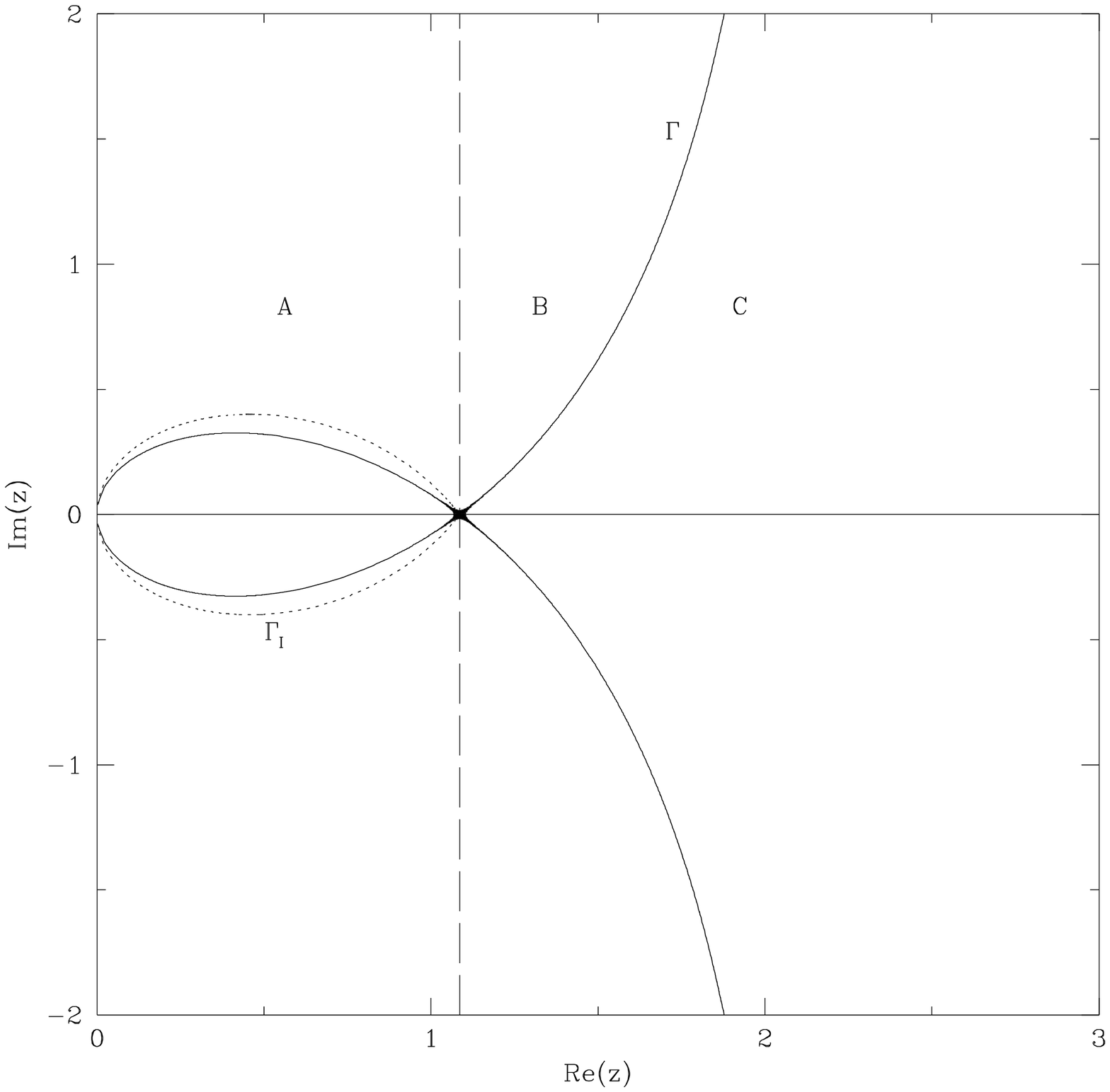,width=6.5in}}
       \caption{Plot of the $z$ plane showing the regions in which the
        possible saddle points are dominant.}
       \label{fig-dom}}
\end{figure}

\begin{figure}
	{\centerline{\psfig{file=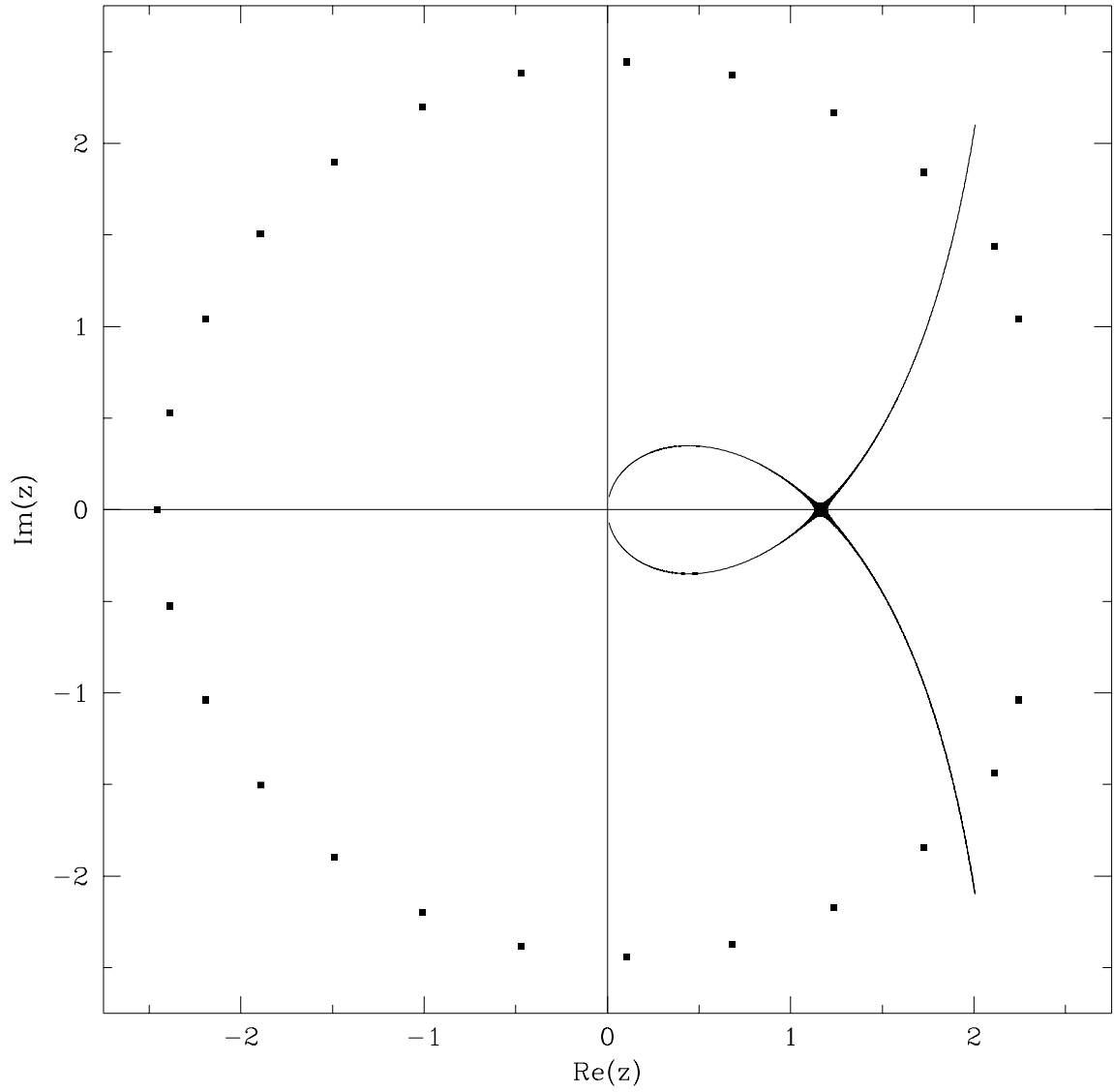,width=6.5in}}
        \caption{Numerical plot of roots of $Z_N(z)=0$ for $\beta=2$,
        $N=25$ and the curve $\Gamma$.}
        \label{fig-root25}}
\end{figure}

\begin{figure}
	{\centerline{\psfig{file=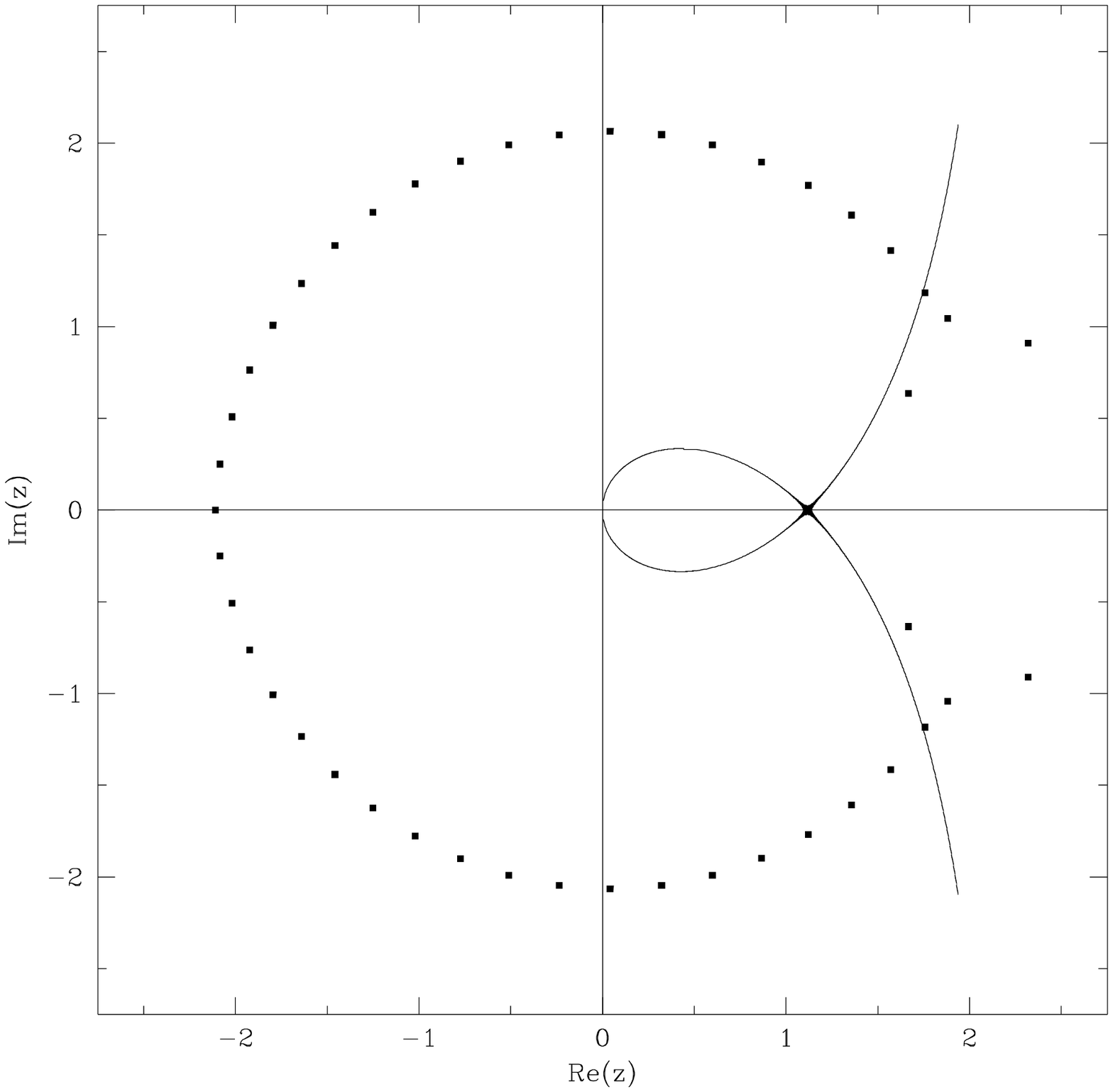,width=6.5in}}
        \caption{As Figure 2, for $N=45$.}
        \label{fig-root45}}
\end{figure}

\begin{figure}
	{\centerline{\psfig{file=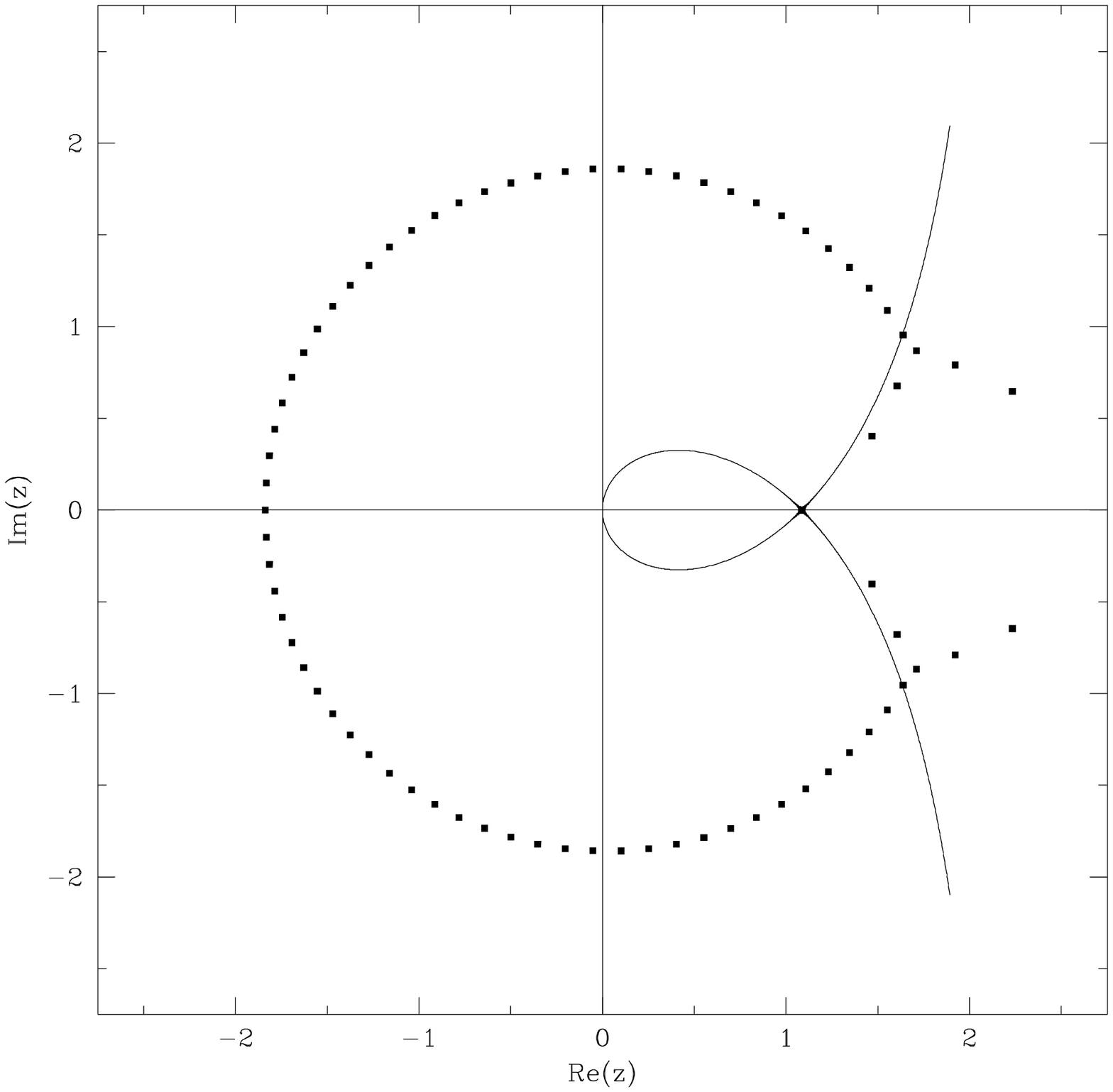,width=6.5in}}
        \caption{As Figure 2, for $N=75$.}
        \label{fig-root75}}
\end{figure}

\end{document}